\documentstyle[preprint,aps,epsf,psfig,axodraw]{revtex}
\tightenlines

\def\dfrac#1#2{\displaystyle\frac{#1}{#2}}
\newcommand{\ovl}[1]{\overline{#1}}
\newcommand{\wt}[1]{\widetilde{#1}}

\newcommand{\p}{\partial}
\newcommand{\kslash}{k\kern-1ex /}
\newcommand{\pslash}{p\kern-1ex /}
\newcommand{\lslash}{l\kern-1ex /}
\newcommand{\sslash}{s\kern-1ex /}
\newcommand{\Dslash}{{\cal D}\kern-1.5ex /}
\newcommand{\bpsi}{\overline{\psi}}

\newcommand{\tr}{{\rm tr}}

\newcommand{\beqa}{\begin{eqnarray}}
\newcommand{\eeqa}{\end{eqnarray}}
\newcommand{\vev}[1]{\left\langle #1 \right\rangle}
\newcommand{\be}{\begin{equation}}
\newcommand{\ee}{\end{equation}}
\newcommand{\ben}{\begin{eqnarray}}
\newcommand{\een}{\end{eqnarray}}
\newcommand{\nn}{\nonumber}
\def\lsim{\raise0.3ex\hbox{$<$\kern-0.75em\raise-1.1ex\hbox{$\sim$}}}
\def\gsim{\raise0.3ex\hbox{$>$\kern-0.75em\raise-1.1ex\hbox{$\sim$}}}
\def\simgt{\rlap{\lower 3.5 pt\hbox{$\mathchar \sim$}}\raise 1pt \hbox {$>$}}
\def\simlt{\rlap{\lower 3.5 pt\hbox{$\mathchar \sim$}}\raise 1pt \hbox {$<$}}

\newcommand{\msbar}{{\overline {\rm MS}}}

\begin{document}
\draft

\title{
\vspace{-5.5cm}
\begin{flushright}
{\normalsize UTHEP-430}\\
\vspace{10mm}
\end{flushright}
Perturbative renormalization factors of 
$\Delta S = 1$ four-quark operators for domain-wall QCD}

\author{Sinya Aoki$^a$ and Yoshinobu Kuramashi$^b$
\thanks{On leave from Institute of Particle and Nuclear Studies,
High Energy Accelerator Research Organization(KEK),
Tsukuba, Ibaraki 305-0801, Japan}}

\address{$^a$Institute of Physics, University of Tsukuba, 
Tsukuba, Ibaraki 305-8571, Japan \\
$^b$Department of Physics, Washington University, 
St. Louis, Missouri 63130, USA \\
}

\date{\today}

\maketitle

\begin{abstract}
Renormalization factors for $\Delta S =1$ four-quark operators 
in the effective weak Hamiltonian are perturbatively
evaluated in domain-wall QCD. The one-loop corrections of 
$\Delta S=1$ four-quark operators consist of two types of
diagrams: one is gluon exchange between quark lines and 
the other is penguin diagrams containing quark loops.
Combining both contributions, our results allow 
a lattice calculation
of the amplitude for $K\rightarrow \pi\pi$ decays with 
$O(g^2)$ corrections included.
\end{abstract}
 
  
\newpage
 

\section{Introduction}

$\Delta S=1$ four-quark operators appear in
the effective low-energy Hamiltonians for non-leptonic weak decays of kaons,
which are relevant for the $\Delta I=1/2$ rule and the $CP$-violation
parameters $\epsilon^\prime/\epsilon$. Although comparison of their
experimental results with  
theoretical predictions based on the standard model is supposed to  
give a good
testing ground for the model, it is not accomplished without
any reliable non-perturbative estimates for the hadronic matrix elements
of $\Delta S=1$ four-quark operators.

Lattice QCD calculation allow us to evaluate 
the $\Delta S=1$ hadronic matrix elements from first
principles in QCD. While in the past decade the Wilson and the 
Kogut-Susskind quark actions are used to calculate the four-quark hadronic
matrix elements, these quark actions have inherent defects: 
the explicit chiral symmetry breaking in the Wilson quark action, which
causes the non-trivial operator mixing among different chiralities, and
the mixture of space-time and flavor symmetries 
in the Kogut-Susskind quark action, which
annoys us by demanding non-trivial matching 
between continuum and lattice operators. 
The domain-wall quark model\cite{shamir,FS} is a five-dimensional Wilson fermion
with free boundaries imposed on the fifth direction of $N_s$ layers.
This quark formulation is expected to 
have superior features in the large $N_s$ limit: no fine tuning
for the chiral limit, $O(a^2)$ scaling violation and no mixing between
four-quark operators with different chiralities.   
Perturbative calculations have shown that these expectations are fulfilled
at the one-loop level in the limit  
$N_s\rightarrow \infty$\cite{pt_at,pt_awi,pt_2,pt_34,NT99}, and
numerical studies support these features 
non-perturbatively\cite{BS,lat98,vpp}.
These advantageous features over other quark formulations urge us to
apply the domain-wall quark to a calculation of 
the $\Delta S=1$ hadronic matrix elements. 

Since the Wilson coefficients in the effective low-energy 
Hamiltonians for non-leptonic weak decays of kaons are calculated
in some continuum renormalization scheme(${\it e.g.}, \msbar$),  
the corresponding  $\Delta S=1$ hadronic matrix elements should
be given in the same continuum scheme, which requires us 
to convert the matrix elements obtained by lattice
simulations to those defined in continuum renormalization 
scheme. A necessary ingredient for this transformation 
is the renormalization factors 
matching the lattice operators 
to the continuum ones.

In this article we make a perturbative calculation of the
renormalization factors for the $\Delta S=1$ four-quark operators
consisting of physical quark fields in the domain-wall QCD(DWQCD).
The one-loop corrections of 
$\Delta S=1$ four-quark operators contain
the gluon exchange diagram and the penguin diagram, which leads to
the following form for the relation between the lattice and continuum operators:
\ben
Q_i^{\rm cont}=\sum_j {{\cal Z}_{ij}^g}Q_j^{\rm latt} 
+ {{\cal Z}_i^{\rm pen}}Q_{\rm pen}^{\rm latt} +O(g^4),
\label{eq:Zfactor}
\een
where $Q_i$ denote a set of $\Delta S=1$ four-quark operators.
The gluon exchange contributions ${{\cal Z}_{ij}^g}$ are already calculated 
in our previous paper\cite{pt_34}, whose
results are applicable to $\Delta S=2$ four-quark operators.
In this work we evaluate the additional contribution denoted by
${{\cal Z}_i^{\rm pen}}Q_{\rm pen}^{\rm latt}$ which originates
from the penguin diagram in the $\Delta S=1$ case.
With the aid of the results in Ref.\cite{pt_34}, we obtain 
complete expressions of the renormalization factors
for the $\Delta S=1$ four-quark operators. 

This paper is organized as follows. In Sec.\ref{sec:penguin} we present
the calculations of the penguin diagram in the continuum and on the lattice.
Full expressions of the renormalization factors for the $\Delta S=1$
operators are given in Sec.\ref{sec:Zfull}.
Our conclusions are summarized in Sec.\ref{sec:conclusion}  
In Appendix we give the DWQCD action and the Feynman rule relevant 
for the present calculation to make this paper self-contained.

The physical quantities are expressed in lattice units
and the lattice spacing $a$ is suppressed unless necessary.
We take SU($N$) gauge
group with the gauge coupling $g$, 
while $N=3$ is specified in the numerical calculations.
Hereafter, repeated indices are to be summed over unless otherwise indicated.

\section{Penguin diagrams}
\label{sec:penguin}
  
\subsection{$\Delta S$=1 four-quark operators}
\label{subsec:op_s1}
  
We consider the following $\Delta S =1$ four-quark operators,
\ben
Q_1 & = &(\bar s_a  u_b)_{V-A}(\bar u_b d_a)_{V-A}, \\
Q_2 & = &(\bar s_a  u_a)_{V-A}(\bar u_b d_b)_{V-A}, \\
Q_3 & = &(\bar s_a  d_a)_{V-A}\sum_q (\bar q_b q_b)_{V-A},\\
Q_4 & = &(\bar s_a  d_b)_{V-A}\sum_q (\bar q_b q_a)_{V-A}, \\
Q_5 & = &(\bar s_a  d_a)_{V-A}\sum_q (\bar q_b q_b)_{V+A},\\
Q_6 & = &(\bar s_a  d_b)_{V-A}\sum_q (\bar q_b q_a)_{V+A}, \\
Q_7 & = &\frac{3}{2}(\bar s_a  d_a)_{V-A}
\sum_q e^q (\bar q_b q_b)_{V+A},\\
Q_8 & = &\frac{3}{2}(\bar s_a  d_b)_{V-A}\sum_q e^q(\bar q_b q_a)_{V+A}, \\
Q_9 & = &\frac{3}{2}(\bar s_a  d_a)_{V-A}
\sum_q e^q (\bar q_b q_b)_{V-A},\\
Q_{10} & = &\frac{3}{2}(\bar s_a  d_b)_{V-A}\sum_q e^q(\bar q_b q_a)_{V-A}, 
\een
where $V\pm A$ indicates the chiral structures $\gamma_\mu(1{\pm}\gamma_5)$. 
$a,b$ denote color indices and $e^q$ are quark charges: $2/3$ for
the up-like quarks and $-1/3$ for the down-like quarks.
The flavor sum is intended over those which are active below 
the renormalization scale of the operators.
This set of operator basis closes under QCD and QED renormalization.
We classify these 10 operators into the following four types:
\ben
Q_1 &=& (\bar s_a  \Gamma_X^\mu u_b )(\bar u_b \Gamma_X^\mu d_a), \\
Q_2 &=& (\bar s_a  \Gamma_X^\mu u_a )(\bar u_b \Gamma_X^\mu d_b), \\
Q_i^{e}&=&(\bar s_a  \Gamma_X^\mu d_b )\sum_q\alpha^q_i
(\bar q_b \Gamma_Y^\mu q_a)  \qquad i=4,6,8,10, \\
Q_i^{o}&=&(\bar s_a  \Gamma_X^\mu d_a )\sum_q\alpha^q_i
(\bar q_b \Gamma_Y^\mu q_b)  \qquad i=3,5,7,9, 
\een
where $\Gamma_X^\mu = \gamma_\mu(1-\gamma_5)$ and 
$\Gamma_Y^\mu = \gamma_\mu(1{\pm}\gamma_5)$. $\alpha^q_i$ are
given by
\ben
\alpha^q_i&=&1\qquad i=3,4,5,6, \\
\alpha^q_i&=&\frac{3}{2}e^q\qquad i=7,8,9,10. 
\een

\subsection{Continuum calculation}
\label{subsec:vf_cont}

It is instructive to first show the calculation of the penguin 
diagram 
in the continuum regularization schemes. For the present calculation
we employ the Naive Dimensional
Regularization(NDR) and the Dimensional Reduction(DRED) 
as the ultraviolet regularization scheme,
in each of which the loop momenta of the Feynman integrals are defined
in $D$ dimensions parameterized by $D=4-2\epsilon$ ($\epsilon>0$). 
The major difference between both regularization schemes are found
in the definitions of the Dirac matrices: NDR scheme defines the Dirac 
matrices in $D$ dimension, while DRED scheme retains them in
four dimensions.

We calculate the following Green function with massless quarks:
\be
\langle Q_i \rangle_{ab;cd}
\equiv \vev{ Q_i s_a(p_1) \bar d_b(p_2)
q_c(p_3) \bar q_d(p_4) },
\label{eq:gf}
\ee
where $a,b,c,d$ label color indices, while
spinor indices are suppressed.
Truncating the external quark propagators from
$\langle Q_i \rangle$, where we multiply $i\pslash_j$ ($j=1,2,3,4$) on 
$\langle Q_i \rangle$, we obtain the vertex functions, 
which is written in the following form up to the
one-loop level
\be
\left[\Lambda_i(p)\right]_{ab;cd}
=\left[\Lambda_i^{(0)}(p)+\Lambda_i^{(1-g)}(p)
+\Lambda_i^{(1-p)}(p)\right]_{ab;cd},
\ee
where the momentum $p$ is defined by $p=p_1-p_2=p_4-p_3$.
The superscript $(i)$ refers to the $i$-th loop level.
At the one-loop level $\Lambda_i^{(1-g)}$ and $\Lambda_i^{(1-p)}$ represent
the gluon exchange diagram and the penguin
diagram, respectively. In this paper, however, we do not consider 
$\Lambda_i^{(1-g)}$ because the gluon exchange contributions to the
renormalization factors are already calculated 
in our previous paper\cite{pt_34}.

The tree-level vertex functions $\Lambda_i^{(0)}$ are given by
\ben
\left[\Lambda_1^{(0)}\right]_{ab;cd}&=&\left[1\wt\odot 1\right]_{ab;cd} 
\Gamma_X^\mu\otimes \Gamma_X^\mu, 
\label{eq:q_1_tree}\\
\left[\Lambda_2^{(0)}\right]_{ab;cd}&=&\left[1\wt\otimes 1\right]_{ab;cd} 
\Gamma_X^\mu\otimes \Gamma_X^\mu, 
\label{eq:q_2_tree}\\
\left[\Lambda_i^{e(0)}\right]_{ab;cd}&=&\alpha_i^u
\left[1\wt\odot 1\right]_{ab;cd} 
\Gamma_X^\mu\otimes \Gamma_Y^\mu, 
\label{eq:q_ie_tree}\\
\left[\Lambda_i^{o(0)}\right]_{ab;cd}&=&\alpha_i^u
\left[1\wt\otimes 1\right]_{ab;cd} 
\Gamma_X^\mu\otimes \Gamma_Y^\mu, 
\label{eq:q_io_tree}
\een
where $\otimes$ represent the tensor structures in the Dirac spinor space and
$\wt{\odot}$ and $\wt{\otimes}$ 
act on the color space as
\ben
\left[1 \wt{\odot} 1\right]_{ab;cd}  & \equiv & \delta_{ad}\delta_{cb}, \\
\left[1 \wt{\otimes} 1\right]_{ab;cd} & \equiv & \delta_{ab}\delta_{cd}.
\een
We take $q=u$ in eq.(\ref{eq:gf}) for convenience.  
This special choice, however,  does not affect the final results
for the renormalization factors. 

The penguin diagram is illustrated in Fig.~\ref{fig:penguin}, which
yields the one-loop level vertex functions $\Lambda_i^{(1-p)}$,
\ben
\left[\Lambda_1^{(1-p)}\right]_{ab;cd} &=& 0, 
\label{eq:vf_c1}\\
\left[\Lambda_2^{(1-p)}\right]_{ab;cd} &=&
-g^2 \left[T^A \wt \otimes T^B\right]_{ab;cd} \ 
G_{\mu\nu}^{AB}(p)\ L_{\alpha\beta}(p)\
\Gamma_X^\delta \gamma_\alpha
\gamma_\mu\gamma_\beta \Gamma_X^\delta \otimes \gamma_\nu, 
\label{eq:vf_c2}\\
\left[\Lambda_i^{e(1-p)}\right]_{ab;cd} &=&
+g^2 \left[T^A \wt \otimes T^B\right]_{ab;cd} \  
G_{\mu\nu}^{AB}(p)\ L_{\alpha\beta}(p)\
\displaystyle \sum_q \alpha_i^q \
\tr( \Gamma_Y^\delta \gamma_\alpha\gamma_\mu\gamma_\beta ) \
\Gamma_X^\delta \otimes \gamma_\nu,\label{eq:vf_c3} \\
\left[\Lambda_i^{o(1-p)}\right]_{ab;cd}  &=& 
-g^2 \left[T^A \wt \otimes T^B\right]_{ab;cd} \ 
G_{\mu\nu}^{AB}(p) \ L_{\alpha\beta}(p)\
\left[ \alpha_i^d\Gamma_X^\delta \gamma_\alpha\gamma_\mu\gamma_\beta 
\Gamma_Y^\delta
+\alpha_i^s \Gamma_Y^\delta \gamma_\alpha\gamma_\mu\gamma_\beta 
\Gamma_X^\delta \right]
\otimes \gamma_\nu, 
\label{eq:vf_c4}
\een
where $T^A$ ($A=1,\dots,N^2-1$) are generators of color SU($N$).
$G_{\mu\nu}^{AB}$ is the gauge propagator given by
\ben
G_{\mu\nu}^{AB}(p)=\left\{
\begin{array}{ll}
\delta_{AB}\delta_{\mu\nu} \dfrac{1}{p^2}  & \qquad \mbox{NDR}, \\
\delta_{AB}\tilde \delta_{\mu\nu} \dfrac{1}{p^2} & \qquad \mbox{DRED}, \\
\end{array} \right. 
\een
with $D$-dimensional metric tensor $\delta_{\mu\nu}$ and
the four-dimensional one $\tilde \delta_{\mu\nu}$.
$L_{\alpha\beta}(p)$ denote the integral of the quark loop, whose
result is 
\ben
L_{\alpha_\beta}(p) &=&\frac{1}{6 (4\pi)^2}
\left[
\left(\frac{1}{\bar\epsilon}+\log(\mu^2/p^2)+\frac{5}{3}+1
\right)\frac{p^2}{2}\delta_{\alpha\beta}
+\left(\frac{1}{\bar\epsilon}+\log(\mu^2/p^2)+\frac{5}{3}
\right)p_\alpha p_\beta
\right],
\een
where $1/\bar\epsilon=1/\epsilon-\gamma+\ln|4\pi|$ and $\mu$ is
the renormalization scale.
We remark that the metric tensor $\delta_{\alpha\beta}$ is
defined in $D$ dimension both for the NDR and DRED schemes.
The products of the Dirac matrices are reduced as follows:
\ben
\tr \left[\Gamma_Y^\nu \gamma_\alpha\gamma_\mu \gamma_\beta 
\right]\delta_{\alpha\beta}
&=& 4\times \left\{
\begin{array}{ll}
(2-D)\delta_{\mu\nu} & \qquad \mbox{NDR}, \\
2\delta_{\mu\nu}-D\tilde\delta_{\mu\nu} & \qquad \mbox{DRED},  \\
\end{array} \right. \\
\tr \left[ \Gamma_Y^\nu \gamma_\alpha\gamma_\mu \gamma_\beta \right]
p_\alpha p_\beta
&=& 4\times \left\{
\begin{array}{ll}
2p_\mu p_\nu - \delta_{\mu\nu} p^2  & \qquad \mbox{NDR}, \\
2p_\mu p_\nu - \tilde\delta_{\mu\nu} p^2  & \qquad \mbox{DRED},  \\
\end{array} \right. \\
\Gamma_X^\nu \gamma_\alpha\gamma_\mu \gamma_\beta\Gamma_Y^\nu 
\delta_{\alpha\beta}
&=& \delta_{XY} \times\left\{
\begin{array}{ll}
2(D-2)^2 \gamma_\mu(1-\gamma_5)  & \qquad \mbox{NDR}, \\
-4\left[2\bar\gamma_\mu-D\gamma_\mu\right](1-\gamma_5) & \qquad \mbox{DRED}, \\
\end{array} \right. \\
\Gamma_X^\nu \gamma_\alpha\gamma_\mu \gamma_\beta\Gamma_Y^\nu p_\alpha p_\beta
&=& \delta_{XY} \times \left\{
\begin{array}{ll}
2(D-2)\left[ p^2\gamma_\mu -2p_\mu \pslash \right](1-\gamma_5) 
 & \qquad \mbox{NDR}, \\
4 \left[p^2\gamma_\mu - 2p_\mu \pslash \right](1-\gamma_5) 
 & \qquad \mbox{DRED}, \\
\end{array}\right.
\een
where $\bar\gamma_\mu = \delta_{\mu\nu}\gamma_\nu$ in the DRED scheme, and
we choose $E_\mu = \bar\gamma_\mu -\frac{D}{4}\gamma_\mu$ as 
an evanescent operator in the DRED scheme\cite{BW}.
It should be noted that the terms proportional to $\pslash$ 
are eliminated by imposing 
on-shell conditions on external quark states.

After some algebra 
we  obtain the following expressions for the vertex functions
(\ref{eq:vf_c1})$-$(\ref{eq:vf_c4}):
\ben
\left[\Lambda_2^{(1-p)}\right]_{ab;cd}& = & -\frac{g^2}{12\pi^2} 
\left[T^A \wt \otimes T^A\right]_{ab;cd} \ 
\gamma_\mu (1-\gamma_5)\otimes\gamma_\mu \ L_i^{\rm cont} \\
\left[\Lambda_i^{e(1-p)}\right]_{ab;cd}& = & -\frac{g^2}{12\pi^2} 
\sum_q\alpha_i^q 
\left[T^A \wt \otimes T^A\right]_{ab;cd} \ 
\gamma_\mu (1-\gamma_5)\otimes\gamma_\mu \ L_i^{\rm cont} \\
\left[\Lambda_i^{o(1-p)}\right]_{ab;cd}& = & -\delta_{XY}\frac{g^2}{12\pi^2} 
(\alpha_i^s+\alpha_i^d)\
\left[T^A \wt \otimes T^A\right]_{ab;cd} \ 
\gamma_\mu (1-\gamma_5)\otimes\gamma_\mu \ L_i^{\rm cont}
\een
where
\ben
L_i^{\rm cont} &=&\frac{1}{\bar\epsilon} 
+ \log( \mu^2/p^2) + \frac{5}{3} +c_i
\een
with 
\ben
c_i = \left\{
\begin{array}{ll}
-1 & \qquad \mbox{$Q_2$ and $Q_i^o$ in NDR}, \\
0  & \qquad \mbox{$Q_i^e$ in NDR}, \\
\frac{1}{4} & \qquad \mbox{$Q_2$, $Q_i^e$ and $Q_i^o$ in DRED}.
\end{array}\right.
\een
The pole term $1/\bar\epsilon$ is subtracted from $L_i^{\rm cont}$
in the $\msbar$ scheme.

From the tree-level vertex functions $\Lambda_i^{(0)}$
in eqs.(\ref{eq:q_1_tree})$-$(\ref{eq:q_io_tree})  
we find
\ben
\left[T^A \wt \otimes T^A\right]_{ab;cd}\ 
\gamma_\mu (1-\gamma_5)\otimes\gamma_\mu & = &
\frac{1}{2}\left [1\wt\odot 1 -\frac{1}{N} 
1\wt\otimes 1\right]_{ab;cd}\
\gamma_\mu (1-\gamma_5)\otimes 
\frac{1}{2}[ \gamma_\mu (1-\gamma_5) + \gamma_\mu (1+\gamma_5) ] \nn \\
&=& \frac{1}{4}\left[ \Lambda_4^{(0)} + \Lambda_6^{(0)} 
- \frac{1}{N}( \Lambda_3^{(0)}+\Lambda_5^{(0)})\right]_{ab;cd}.
\een
Using this relation we finally obtain
\ben
\Lambda_i^{(1-p)} &=& -\frac{g^2}{12\pi^2}\frac{1}{4}\ C(Q_i)\ L_i^{\rm cont}
\left[ \Lambda_4^{(0)} + \Lambda_6^{(0)} 
- \frac{1}{N}( \Lambda_3^{(0)}+\Lambda_5^{(0)})\right]
\label{eq:o_pen_cont}
\een
with $C(Q_2)$=1, $C(Q_3)=2$, $C(Q_4)=C(Q_6)=f_q$,
$C(Q_8)=f_u-f_d/2$, $C(Q_9)=-1$, $C(Q_{10})=f_u-f_d/2$ 
and $C(Q_i)=0$ for other $i$, where $f_q$, $f_u$ and $f_d$ denote the
number of flavors, the number of charge 2/3 up-like quarks and
the number of charge $-1/3$ down-like quarks, respectively.

\subsection{Lattice calculation}
\label{subsec:vf_latt}

Let us turn to the calculation of the vertex functions on the lattice.
In this subsection we use the same notations for quantities defined on the
lattice as those for their counterparts in the continuum. 

We consider the Green function of eq.~(\ref{eq:gf}) on the lattice.
At the one-loop level the penguin diagram in Fig.~\ref{fig:penguin}
contributes to the vertex functions:
\ben
\left[\Lambda_1^{(1-p)}\right]_{ab;cd} &=& 0,
\label{eq:vf_l1}\\                       
\left[\Lambda_2^{(1-p)}\right]_{ab;cd} &=&   
-g^2 \left[T^A \wt \otimes T^B\right]_{ab;cd} \ 
 (1-w_0^2)^3 G_{\mu\nu}^{AB}(p) 
\Gamma_X^\delta L_\mu (p) \Gamma_X^\delta \otimes \gamma_\nu, 
\label{eq:vf_l2}\\     
\left[\Lambda_i^{e(1-p)}\right]_{ab;cd} &=&                       
+g^2 \left[T^A \wt \otimes T^B\right]_{ab;cd} \ 
(1-w_0^2)^3 G_{\mu\nu}^{AB}(p)
\sum_q \alpha^q_i
\tr( \Gamma_Y^\delta  L_\mu(p) ) \Gamma_X^\delta\otimes \gamma_\nu,
\label{eq:vf_le} \\
\left[\Lambda_i^{o(1-p)}\right]_{ab;cd}  &=&      
-g^2 \left[T^A \wt \otimes T^B\right]_{ab;cd} \ 
(1-w_0^2)^3 G_{\mu\nu}^{AB}(p)
\left[
\alpha^d_i\Gamma_X^\delta L_\mu (p)\Gamma_Y^\delta +
\alpha^s_i\Gamma_Y^\delta L_\mu (p)\Gamma_X^\delta
\right]\otimes \gamma_\nu,
\label{eq:vf_lo} 
\een
where $(1-w_0^2)^3$ is the overlap factor to the four-dimensional quark fields
which emerges through the truncation of the external quark legs by
multiplying $i\pslash_i$ ($i=1,\dots,4$) on the Green function.
$L_\mu$ denotes the integral of the quark loop, which is expressed by
\ben
L_\mu (p) &=& \int_{-\pi}^{\pi}\frac{d^4 k}{(2\pi)^4}    
\{ P_R S_F(k)_{1s}+P_L S_F(k)_{N_s s}\} (-i)v_\mu(k-p/2)_{st} \nn\\
& & \times \{ S_F(k-p)_{t1}P_L + S_F(k-p)_{tN_s}P_R\}\\
&=& \int_{-\pi}^{\pi}\frac{d^4 k}{(2\pi)^4} I_\mu(k,p),
\een
with
\ben
I_\mu(k,p)&=&    
-c_\mu(k-p/2)\frac{1}{1-e^{(\alpha+\alpha')}}\gamma_\mu
-c_\mu(k-p/2)\frac{1}{F F'(1-e^{-(\alpha+\alpha')})} 
\sslash \gamma_\mu \sslash' \nn\\
&&+ir s_\mu (k-p/2)\frac{e^{-\alpha}}
{F'(1-e^{-(\alpha+\alpha')})}i\sslash' 
+ir s_\mu (k-p/2) \frac{e^{-\alpha'}}
{F(1-e^{-(\alpha+\alpha')})}i\sslash.
\een
where $c_\mu(k)={\rm cos}k_\mu$, $s_\mu(k)={\rm sin}k_\mu$, $s'_\mu=s_\mu(k-p)$
$\alpha = \alpha (k)$, $\alpha' = \alpha (k-p)$,
$F=F(k)$ and  $F' =F(k-p)$.
The expressions of $v_\mu$, $S_F$, $\alpha$ and $F$ are given in Appendix.

Since the function $L_\mu(p)$ has an infrared divergence
for $p^2\rightarrow 0$, we consider extracting its divergent part
by employing an analytically integrable expression 
$\tilde I_\mu(k,p)$ which has the same infrared behavior as $I_\mu(k,p)$, 
\ben
L_\mu (p) &=& \int_{-\pi}^{\pi}\frac{d^4 k}{(2\pi)^4}\tilde I_\mu(k,p) 
+ \int_{-\pi}^{\pi}\frac{d^4 k}{(2\pi)^4} 
\left\{I_\mu(k,p)-\tilde I_\mu(k,p)\right\}, 
\label{eq:Lsubtaction}
\een
where we choose
\be
\tilde I_\mu(k,p)=\theta(\Lambda^2-k^2)(1-w_0^2)\frac{1}{i\kslash}\gamma_\mu
\frac{1}{i(\kslash-\pslash)},
\ee
with $\Lambda$ $(\le\pi)$ a cut-off.
Now the second term in the right hand side of eq.(\ref{eq:Lsubtaction}) is
regular in terms of $p$, we can make a Taylor expansion,
\ben
L_\mu (p) = \int_{-\pi}^{\pi}\frac{d^4 k}{(2\pi)^4} \tilde I_\mu(k,p) 
&+&\int_{-\pi}^{\pi}\frac{d^4 k}{(2\pi)^4} 
\left\{I_\mu(k,0)-\tilde I_\mu(k,0)\right\} \nn\\ 
&+& \int_{-\pi}^{\pi}\frac{d^4 k}{(2\pi)^4}
p_\rho \frac{\p}{\p p_\rho}
\left\{I_\mu(k,p)-\tilde I_\mu(k,p)\right\}\vert_{p=0} \nn\\
&+& \int_{-\pi}^{\pi}\frac{d^4 k}{(2\pi)^4}
\frac{1}{2} p_\rho p_\lambda \frac{\p^2}{\p p_\rho \p p_\lambda}
\left\{I_\mu(k,p)-\tilde I_\mu(k,p)\right\}\vert_{p=0},
\label{eq:Lexpansion}
\een
where $\rho,\lambda=1,2,3,4$.
Before evaluating each term of the right hand side, 
let us consider the transformation properties of $L_\mu$ under various
discrete symmetries, from which we can predict the types of terms
to be allowed in $L_\mu$. The same kind of discussion is found 
in the Wilson quark case\cite{penguin_w}.
    
Under the operation of charge conjugation matrix $C=\gamma_4\gamma_2$ we find
\ben
C v_\mu(k)_{st} C^{-1}&=&-v_\mu^T(-k)_{N_s+1-t,N_s+1-s}, \\
C S_F(k)_{st} C^{-1}&=& S_F^T(-k)_{N_s+1-t,N_s+1-s}, \\
C L_\mu(p)  C^{-1}&=&-L_\mu^T(p).
\een
Parity transformation gives
\ben
\gamma_4 v_4(k_4,{\vec k})_{st} \gamma_4&=&  
v_4(k_4,-{\vec k})_{N_s+1-s,N_s+1-t}, \\
\gamma_4 v_i(k_4,{\vec k})_{st} \gamma_4&=&  
-v_i(k_4,-{\vec k})_{N_s+1-s,N_s+1-t}, \\
\gamma_4 S_F(k_4,{\vec k})_{st} \gamma_4&=& 
S_F(k_4,-{\vec k})_{N_s+1-s,N_s+1-t}, \\
\gamma_4 L_4(p_4,{\vec p})  \gamma_4&=&L_4(p_4,-{\vec p}),\\
\gamma_4 L_i(p_4,{\vec p})  \gamma_4&=&-L_i(p_4,-{\vec p}),
\een
where $i=1,2,3$.
These discrete symmetries  restrict the form of $L_\mu$ as
\be
L_\mu(p)=\frac{1}{a^2}c_0\gamma_\mu+\frac{1}{a}c_1 i\sigma_{\mu\nu}p_\nu
+c_{2a}\gamma_\mu p^2+c_{2b} p_\mu\pslash
+c_{2c}\gamma_\mu p_\mu^2 +O(a),
\label{eq:Lform}
\ee
with $\sigma_{\mu\nu}=[\gamma_\mu,\gamma_\nu]/2$, where 
the massless quark case is considered.
We can eliminate further terms by the Ward-Takahashi identity:
\be
2{\rm sin}(p_\mu a/2)v_\mu(k-p/2)_{st}=
[S_F(k)^{-1}]_{st}-[S_F(k-p)^{-1}]_{st},
\ee
which leads to
\ben
2{\rm sin}(p_\mu a/2)L_\mu(p)&=&i\int_{-\pi}^{\pi}\frac{d^4 k}{(2\pi)^4}
\left[P_R\left\{S_F(k)_{11}-S_F(k-p)_{11}\right\}P_L \right.\nn\\
&&\qquad +P_R\left\{S_F(k)_{1N_s}-S_F(k-p)_{1N_s}\right\}P_R \nn\\
&&\qquad +P_L\left\{S_F(k)_{N_s 1}-S_F(k-p)_{N_s 1}\right\}P_L \nn\\
&&\left.\qquad +P_L\left\{S_F(k)_{N_s N_s}-S_F(k-p)_{N_s N_s}\right\}P_R\right] \nn\\
&=&0
\een
where the periodicity of $S_F$ is used. 
Under the requirement that the terms of the left hand side must vanish
order by order in terms of the lattice spacing,  we find
$c_0=c_{2c}=0$ and $c_{2a}=-c_{2b}$ in eq.(\ref{eq:Lform}), with which
the expression of $L_\mu$ is simplified as
\be
L_\mu(p)=\frac{1}{a}c_1 i\sigma_{\mu\nu}p_\nu
+c_{2a}(\gamma_\mu p^2-p_\mu\pslash)
+O(a).
\label{eq:Lform_wi}
\ee
Here we should note that 
\ben
&&\gamma_\delta \sigma_{\mu\nu} \gamma_\delta=0, \\
&&\tr\left(\sigma_{\mu\nu}\gamma_\delta(1\pm\gamma_5)\right)=0,
\een
which means $\sigma_{\mu\nu}p_\nu$ term in eq.(\ref{eq:Lform_wi})
does not contribute to the wave functions (\ref{eq:vf_l1})$-$(\ref{eq:vf_lo}).

Now we see that $L_\mu(0)=0$ and linear term in $p$ is irrelevant
in the expansion (\ref{eq:Lexpansion}), we focus on the remaining terms. 
Performing the integration of $\tilde I_\mu(k,p)-\tilde I_\mu(k,0)$ analytically,
we obtain
\ben
\int_{-\pi}^{\pi}\frac{d^4 k}{(2\pi)^4}
\left\{\tilde I_\mu (k,p)-\tilde I_\mu(k,0)\right\} 
&=& (1-w_0^2)\frac{1}{16\pi^2}
\left[ \left(-p^2 \gamma_\mu + p_\mu \pslash \right)\frac{1}{3}
\left( \log (\Lambda^2/p^2) + 5/6 \right) - \frac{p^2}{6}\gamma_\mu
\right].
\label{eq:IRterm}
\een
As for the last term in the expansion of eq.~(\ref{eq:Lexpansion}), 
some tedious  algebra leads to the following expression for the integrand:
\beqa
&&p_\rho p_\lambda \frac{\p^2}{\p p_\rho \p p_\lambda}
\left\{I_\mu(p)-\tilde I_\mu(p)\right\}\vert_{p=0}  \nn \\
&&=p^2 \gamma_\mu \left( R_{\mu\not= \nu}^b - R^d_{\mu\not= \nu\nu}\right)
+  p_\mu \pslash \left( 2 R_{\mu\not= \nu}^c + R^d_{\mu\mu\not=\nu}
+R^d_{\mu\not=\nu\mu}\right) \nn\\
&&+p_\mu^2\gamma_\mu \left(R_\mu^a + R_{\mu\mu}^b-R_{\mu\not=\nu}^b
+2 R_{\mu\mu}^c-2 R_{\mu\not=\nu}^c 
+R^d_{\mu\mu\mu}+R^d_{\mu\not= \nu\nu} - R^d_{\mu\mu\not=\nu}
-R^d_{\mu\not=\nu\mu} \right) \nn\\
&&+ \gamma_\nu\gamma_\alpha p_\nu p_\alpha \gamma_\mu
(R_{\mu\not=\alpha\not=\nu}^d-R_{\mu\not=\nu\not=\alpha}^d),
\label{eq:p2term}
\eeqa
where $R_\mu^a$, $R_{\mu\nu}^b$, $R_{\mu\nu}^c$ and $R_{\mu\nu\alpha}^d$
are given  by 
\ben
R_\mu^a &=& -e^{-\alpha}\frac{c_\mu}{4(e^{\alpha}-e^{-\alpha})}
+\frac{s_\mu^2 f_\mu}{(e^\alpha-e^{-\alpha})^2} \nn \\
&&+\frac{c_\mu( 10 s_\mu^2 -s^2)}{4 F^2(1-e^{-2\alpha})} 
-\frac{s_\mu^2 s^2 F_\mu}{F}
-\frac{4 c_\mu s_\mu^4 G_{\mu\mu}}{F} \nn\\
&&+r\frac{5 s_\mu^2}{4 F(e^\alpha-e^{-\alpha})}
-2r e^{-\alpha} s_\mu^4 G_{\mu\mu}  \nn \\
&&+\frac{rs_\mu^2}{4 F (e^\alpha-e^{-\alpha})} +
\frac{2r s_\mu^4 e^\alpha}{F(e^\alpha-e^{-\alpha})^3}
\left[
f_\mu^2 (e^\alpha+e^{-\alpha})-g_{\mu\mu}(e^\alpha-e^{-\alpha}) 
\right] \nn \\
&&- 32 (1-w_0^2)\theta (\Lambda^2-k^2)\frac{k_\mu^4}{ (k^2)^4},  \\
R_{\mu\nu}^b &=&
\frac{c_\mu}{(e^\alpha-e^{-\alpha})^3}
\left[
s_\nu^2f_\nu^2 (e^\alpha+e^{-\alpha})
-(h_\nu+s_\nu^2 g_{\nu\nu})(e^\alpha-e^{-\alpha}) 
\right] \nn \\
&&-\frac{c_\mu s_\nu^2}{F^2(1-e^{-2\alpha})} 
+c_\mu\frac{(2s_\mu^2-s^2)(H_\nu+ s_\nu^2G_{\nu\nu})}{F} \nn\\
&&+ r e^{-\alpha} s_\mu^2 (s_\nu^2 G_{\nu\nu}+H_\nu) \nn\\
&&+ r e^{\alpha}\frac{s_\mu^2}{F(e^\alpha-e^{-\alpha})^3}
\left[
(s_\nu^2g_{\nu\nu}+h_\nu)(e^\alpha-e^{-\alpha})-
s_\nu^2 f_\nu^2 (e^\alpha+e^{-\alpha})
\right] \nn\\
&&-(1-w_0^2)\theta (\Lambda^2-k^2)\frac{ 2 (k^2-2 k_\mu^2)(4 k_\nu^2 - k^2) }
{(k^2)^4}, \\
R_{\mu\nu}^c &=& e^\alpha \frac{c_\nu s_\mu^2 }{2F^2(e^\alpha-e^{-\alpha})} 
+ s_\nu^2 F_\nu\frac{ s_\mu^2}{F} 
+\frac{ 2 c_\mu s_\mu^2 s_\nu^2 G_{\mu\nu}}{F} \nn \\
&&- r \frac{ c_\mu c_\nu }{ 2 F (e^\alpha-e^{-\alpha})}
+ r e^{-\alpha} \left[
s_\mu^2 s_\nu^2 G_{\mu\nu} - s_\mu^2 c_\nu F_\mu 
- c_\mu s_\nu^2 F_\nu/2
\right] \nn \\
&&+r e^{\alpha}\frac{c_\mu s_\nu^2 f_\nu}{ 2 F (e^\alpha-e^{-\alpha})^2}
+r e^{\alpha} \frac{s_\mu^2 s_\nu^2}{ F (e^\alpha-e^{-\alpha})^3}
\left[
g_{\mu\nu} (e^\alpha-e^{-\alpha}) -f_\mu f_\nu (e^\alpha +e^{-\alpha}) 
\right] \nn \\
&&+(1-w_0^2)\theta (\Lambda^2-k^2) 16 k_\nu^2 \frac{  k_\mu^2}{(k^2)^4}, \\
R_{\mu\nu\alpha}^d &=& -2 \frac{c_\mu s_\nu^2 F_\nu c_\alpha}{F}
- 4 (1-w_0^2)\theta (\Lambda^2-k^2)\frac{k_\nu^2}{(k^2)^3},
\een
with
\ben
f_\mu &=&\frac{-r W + c_\mu + r\cosh\alpha}{W\sinh\alpha}, \\
h_\mu &=&\frac{-c_\mu r W + c_\mu^2-s_\mu^2 + rc_\mu\cosh\alpha}
{W\sinh\alpha}, \\
g_{\mu\nu} &=&  g_{\nu\mu}= -\frac{f_\mu f_\nu}{\tanh\alpha}+
\frac{r^2}{W\sinh\alpha}+r\frac{f_\mu+f_\nu}{W},  \\
F_\mu &=& e^{2\alpha}\frac{-r+W f_\mu}{F^2(e^\alpha-e^{-\alpha})}
-\frac{f_\mu}{F(e^\alpha-e^{-\alpha})^2}, \\
H_\mu &=& e^{2\alpha}\frac{ rc_\mu-Wh_\mu}{F^2(e^\alpha-e^{-\alpha})}
+\frac{h_\mu}{F(e^\alpha-e^{-\alpha})^2}, \\
G_{\mu\nu}&=& G_{\nu\mu}=\frac{1}{F}
\left[\frac{g_{\mu\nu}}{(e^\alpha-e^{-\alpha})^2}-
f_\mu f_\nu\frac{e^\alpha+e^{-\alpha}}{(e^\alpha-e^{-\alpha})^3}
\right] +e^\alpha\frac{f_\mu(-r+Wf_\nu)+f_\nu(-r+Wf_\mu)}
{F^2(e^\alpha-e^{-\alpha})^2} \nn \\
&-&\frac{e^\alpha}{F^3(e^\alpha-e^{-\alpha})}\left[
2e^{2\alpha}(-r+Wf_\mu)(-r+Wf_\nu)+e^\alpha F \{
-r(f_\mu+f_\nu)+Wf_\mu f_\nu + Wg_{\mu\nu}\}
\right].
\een
We choose $r=-1$ in this calculation.
Using the results of eqs.(\ref{eq:IRterm}) 
and (\ref{eq:p2term}), $L_\mu (p)$ is expressed as
\beqa
L_\mu (p) &=& \frac{(1-w_0^2)}{16\pi^2}
\left[ \left(p_\mu \pslash - p^2 \gamma_\mu \right)\frac{1}{3}
\left( \log (\Lambda^2/p^2) + 5/6 \right) \right.\nn \\
&&\left.+ A p^2 \gamma_\mu + B p_\mu \pslash + \Delta R p_\mu^2 \gamma_\mu
+ \delta R \gamma_\nu\gamma_\alpha p_\nu p_\alpha \gamma_\mu
\right],
\eeqa
where
\ben
A &=& -\frac{1}{6}+\frac{16\pi^2}{1-w_0^2}
\frac{1}{2}\int_{-\pi}^{\pi}\frac{d^4 k}{(2\pi)^4}
\left\{R_{\mu\nu}^b-R_{\mu\nu\nu}^d
\right\}, \\
B &=& \frac{16\pi^2}{1-w_0^2}\frac{1}{2}\int_{-\pi}^{\pi}\frac{d^4 k}{(2\pi)^4}
\left\{2 R_{\mu\nu}^c+R_{\mu\mu\nu}^d
+R_{\mu\nu\mu}^d \right\}, \\
\Delta R &=& \frac{16\pi^2}{1-w_0^2}\frac{1}{2}
\int_{-\pi}^{\pi}\frac{d^4 k}{(2\pi)^4}\left\{
R_\mu^a+R_{\mu\mu}^b-R_{\mu\nu}^b+2(R_{\mu\mu}^c-R_{\mu\nu}^c)
+ R_{\mu\mu\mu}^d+R_{\mu\nu\nu}^d-R_{\mu\mu\nu}^d-R_{\mu\nu\mu}^d \right\}, \\
\delta R &=& \frac{16\pi^2}{1-w_0^2}\frac{1}{2}
\int_{-\pi}^{\pi}\frac{d^4 k}{(2\pi)^4}\left\{
R_{\mu\alpha\nu}^d-R_{\mu\nu\alpha}^d\right\}=0, 
\eeqa
with $\mu \not= \nu\not=\alpha$. We find that $\delta R=0$ 
from the symmetry of the integrand.
The integrals are estimated 
by a mode sum for a periodic box of a size $L^4$ with $L=64$ 
after transforming
the momentum variable through $p_\mu=q_\mu-\sin q_\mu$.
We choose $\Lambda=\pi$ for the cut-off in the integrand $\tilde I_\mu$.  
The numerical results for $A$ are $B$ are presented in Table~\ref{tab:penguin}
as a function of $M$. We observe that the expected relation $A=-B$ 
is well satisfied. We also checked that $\Delta R$ is consistent with
zero as expected.
Finally we obtain
\ben
L_\mu (p) &=& \frac{(1-w_0^2)}{16\pi^2}\frac{L^{\rm latt}}{3}
\left[p_\mu\pslash - p^2\gamma_\mu \right],
\een
where
\ben
L^{\rm latt} &=& \log (\pi^2/p^2)+\frac{5}{6}+3B.
\een

Substituting the above expression for $L_\mu(p)$ 
in eqs.(\ref{eq:vf_l1})$-$(\ref{eq:vf_lo}),
we have
\beqa
\left[\Lambda_1^{(1-p)}\right]_{ab;cd}  &=& 0, \\
\left[\Lambda_2^{(1-p)}\right]_{ab;cd}  &=& 
-\frac{g^2}{12\pi^2} (1-w_0^2)^4 L^{\rm latt}
\left[T^A \wt \otimes T^A\right]_{ab;cd} \ 
\gamma_\mu (1-\gamma_5)\otimes\gamma_\mu, \\
\left[\Lambda_i^{e(1-p)}\right]_{ab;cd}  &=& -\frac{g^2}{12\pi^2} 
(1-w_0^2)^4\sum_q\alpha_i^q L^{\rm latt}
\left[T^A \wt \otimes T^A\right]_{ab;cd} \ 
\gamma_\mu (1-\gamma_5)\otimes\gamma_\mu, \\
\left[\Lambda_i^{o(1-p)}\right]_{ab;cd}  &=& -\delta_{XY}\frac{g^2}{12\pi^2} 
(1-w_0^2)^4 (\alpha_i^s+\alpha_i^d) L^{\rm latt}
\left[T^A \wt \otimes T^A\right]_{ab;cd} \ 
\gamma_\mu (1-\gamma_5)\otimes\gamma_\mu,
\eeqa
where we used the on-shell conditions for the external quark states
to eliminate the terms proportional to $\pslash$.
In the same way to derive eq.(\ref{eq:o_pen_cont}) in the continuum calculation,
the lattice wave functions are written in the compact form as follows,  
\ben
\Lambda_i^{(1-p)} &=& -\frac{g^2}{12\pi^2}\frac{1}{4}\ C(Q_i)\ L^{\rm latt}
\left[ \Lambda_4^{(0)} + \Lambda_6^{(0)}
- \frac{1}{N}( \Lambda_3^{(0)}+\Lambda_5^{(0)})\right],
\label{eq:o_pen_latt}
\een
where $C(Q_i)$ are already obtained in the previous subsection.

Comparing the continuum vertex functions in eq.~(\ref{eq:o_pen_cont}) 
and lattice ones in eq.~(\ref{eq:o_pen_latt}),
we find that the penguin diagram contributions to the renormalization factor
in eq.(\ref{eq:Zfactor}) are written as
\be
{{\cal Z}_i^{\rm pen}}Q_{\rm pen}^{\rm latt}=\frac{1}{(1-w_0^2)^2}
Z_i^{\rm pen} \ [Q_4+Q_6-\frac{1}{N}(Q_3+Q_5)]^{\rm latt},
\ee
where the penguin operator and its coefficient are given by
\ben
Q_{\rm pen}^{\rm latt} &=& 
\left[Q_4+Q_6-\dfrac{1}{N}\left( Q_3+Q_5\right)\right]^{\rm latt} 
\label{eq:Q_pen}\\
Z_i^{\rm pen}&=&\frac{g^2}{12\pi^2}\frac{C(Q_i)}{4}
(L^{\rm latt}-L_i^{\rm cont})=
\dfrac{g^2}{16\pi^2}\dfrac{C(Q_i)}{3}
\left[ -\log (\mu a)^2 + z_i^{\rm pen} \right] 
\label{eq:Z_pen}
\een
with $z_i^{\rm pen} = \log(\pi^2)+ 3 B -\dfrac{5}{6}-c_i$.
Numerical values of $z_i^{\rm pen}$ with the DRED scheme are presented in
Table~\ref{tab:penguin} as a function of $M$.

\section{Full renormalization factors}
\label{sec:Zfull}


In order to write down the complete expressions for the 
renormalization factors of the $\Delta S=1$
operators $Q_i$, we still need to know the
contributions from the gluon exchange diagrams, which is
denoted by ${{\cal Z}_{ij}^g}Q_j^{\rm latt}$ 
in eq.~(\ref{eq:Zfactor}). Fortunately, they can be
obtained from the combinations of the results in our previous paper\cite{pt_34},
where we calculated the gluon exchange diagrams for the following 
four-quark operators,
\ben
{\cal O}_\pm (q_1,q_2,q_3,q_4)& = & \frac{1}{8} \left[
(\bar q_1 q_2)_{V-A}(\bar q_3 q_4)_{V-A}
\pm
(\bar q_1 q_4)_{V-A}(\bar q_3 q_2)_{V-A} \right], \\
{\cal O}_1 (q_1,q_2,q_3,q_4)& = &
\frac{1}{4} \left[-C_F (\bar q_1 q_2)_{V-A}(\bar q_3 q_4)_{V+A}
+ (\bar q_1 T^A q_2)_{V-A}(\bar q_3 T^A q_4)_{V+A} \right], \\
{\cal O}_2 (q_1,q_2,q_3,q_4)& = &
\frac{1}{4} \left[ \frac{1}{2N} (\bar q_1  q_2)_{V-A}(\bar q_3 q_4)_{V+A}
+ (\bar q_1 T^A  q_2)_{V-A}(\bar q_3 T^A q_4)_{V+A} \right],
\een
with $C_F=(N^2-1)/(2N)$ the second Casimir of SU($N$) group. 
The tensor structures in the color space are $1\wt\otimes 1$ and 
$T^A \wt\otimes T^A$.

The operators $Q_i$ are related to 
${\cal O}_{\pm,1,2}$ as
\ben
Q_2 \pm Q_1 & = & 8 {\cal O}_\pm (s,u,u,d), \\
Q_3\pm Q_4 & = &  8 \sum_q {\cal O}_\pm (s,d,q,q), \\
Q_9\pm Q_{10} & = &  8 \sum_q \frac{3 e^q}{2}{\cal O}_\pm (s,d,q,q), \\
Q_6 &=& 8 \sum_q {\cal O}_2 (s,d,q,q), \\
Q_8 &=& 8 \sum_q \frac{3 e^q}{2}{\cal O}_2 (s,d,q,q), \\
-N Q_5 + Q_6 &=& 8 \sum_q {\cal O}_1 (s,d,q,q), \\
-N Q_7 + Q_8 &=& 8 \sum_q \frac{3 e^q}{2}{\cal O}_1 (s,d,q,q).
\een

From these relations
the renormalized operators, which include the contributions of
both the gluon exchange and  penguin diagrams, are expressed 
in terms of the lattice operators as follows:
\ben
Q_i^{\rm cont} &=& \dfrac{1}{(1-w_0^2)^2 Z_w^2}
\left[ Z^g_{ij} Q_j^{\rm latt} + Z^{\rm pen}_i Q_{\rm pen}^{\rm latt} \right]
\een
where
\ben
Z^g_{ii} &=& \left\{
\begin{array}{ll}
1+\dfrac{g^2}{16\pi^2}\left[
\dfrac{3}{N}\log(\mu a)^2 +\dfrac{z_+ + z_-}{2}\right]
&\quad i=1,2,3,4,9,10, \\
1+\dfrac{g^2}{16\pi^2}\left[
-\dfrac{3}{N}\log(\mu a)^2 +z_1-v_{21}\right]
&\quad i=5,7, \\
1+\dfrac{g^2}{16\pi^2}\left[
\dfrac{3(N^2-1)}{N}\log(\mu a)^2 +z_2+v_{21}\right]
&\quad i=6,8, \\
\end{array}\right. 
\een
\ben
Z^g_{ij} &=& \left\{
\begin{array}{ll}
= Z^g_{ji}=\dfrac{g^2}{16\pi^2}
\left[-3 \log(\mu a)^2 +\dfrac{z_+-z_-}{2}\right] 
&\quad (i,j) =(1,2),(3,4),(9,10), \\
\dfrac{g^2}{16\pi^2}
\left[3 \log(\mu a)^2 +\dfrac{z_2-z_1+v_{21}-v_{12}}{N}\right]
&\quad (i,j) =(5,6),(7,8), \\
-\dfrac{g^2}{16\pi^2} N v_{21}
&\quad (i,j) =(6,5),(8,7), \\
0 &\quad\mbox{others} \\
\end{array}\right. 
\een
with $z_{\pm,1,2}$,  $v_{12}$, and $v_{21}$ presented 
in our previous paper\cite{pt_34}.
The penguin operator $Q^{\rm latt}_{\rm pen}$ and 
its coefficient $Z_i^{\rm pen}$ are given in eqs.~(\ref{eq:Q_pen})
and (\ref{eq:Z_pen}), respectively. 
For the sake of convenience, 
we list the differences between the NDR and DRED schemes 
in the finite part of the renormalization constant,
\ben
\left(\dfrac{z_+ + z_-}{2}\right)^{\rm NDR} &=& 
\left(\dfrac{z_+ + z_-}{2}\right)^{\rm DRED}
-\dfrac{N^2-6}{2N}, \\
(z_1-v_{21})^{\rm NDR} &=&
(z_1-v_{21})^{\rm DRED} -\dfrac{N^2-8}{2N},\\
(z_2+v_{21})^{\rm NDR} &=&
(z_2+v_{21})^{\rm DRED} -\dfrac{N^2-4}{N},\\
\left(\dfrac{z_+ - z_-}{2}\right)^{\rm NDR} &=& 
\left(\dfrac{z_+ - z_-}{2}\right)^{\rm DRED}
-\dfrac{5}{2}, \\
\left(\dfrac{z_2-z_1+v_{21}-v_{12}}{N}\right)^{\rm NDR} &=&
\left(\dfrac{z_2-z_1+v_{21}-v_{12}}{N}\right)^{\rm DRED} - \dfrac{7}{2}, \\
-N (v_{21})^{\rm NDR} &=&-N (v_{21})^{\rm DRED} -3, \\
(z_i^{\rm pen})^{\rm NDR} &=& (z_i^{\rm pen})^{\rm DRED} +
\left\{
\begin{array}{ll}
\dfrac{5}{4} &\quad  i = 2,3,5,7,9, \\
\dfrac{1}{4} &\quad  i = 4,6,8,10, \\
\end{array}\right.
\een
where $v_{12}=v_{21}=0$ in the DRED scheme.

We are interested in the magnitude of the renormalization factors with the
currently accessible coupling constant in numerical simulations.
Let us estimate it taking $\beta=6.0$ with $M=1.8$  
as a representative case. 
All the necessary ingredients in this analysis 
are given  in our previous papers\cite{pt_2,pt_34} 
and Table~\ref{tab:penguin} of this report.
With the use of the mean field improvement\cite{MF}, we have
\ben
&&(1-w_0(\tilde M)^2) Z_w^{\rm MF}(\tilde M) = 0.9085, \\
&&\frac{z_+^{\rm MF}(\tilde M)+z_-^{\rm MF}(\tilde M)}{2} = -12.848, \\ 
&&\frac{z_+^{\rm MF}(\tilde M)-z_-^{\rm MF}(\tilde M)}{2} = 2.491, \\ 
&&z_1^{\rm MF}(\tilde M)=-11.187, \\
&&z_2^{\rm MF}(\tilde M)=-18.6, \\
&&z_2^{\rm MF}(\tilde M)-z_1^{\rm MF}(\tilde M)=-7.4, \\
&&z_i^{\rm pen}(\tilde M) = 2.328,
\een
in the DRED scheme, where we employ
\ben
g^2_{\ovl{MS}}(1/a) & = &
\left[ P \dfrac{\beta}{6} - 0.13486 \right]^{-1}
= 2.1792, \\
\tilde M &=& M + 4(u-1) = 1.3112,
\een
with $P=u^4=0.59374$ the plaquette value at $\beta=6.0$ 
in the quenched approximation.
Combining these results we obtain
\ben
\left(\dfrac{Z^g_{ii}}{(1-w_0^2)^2 Z_w^2}\right)^{\rm MF} &=& \left\{
\begin{array}{ll}
u^2\times 0.9968 &\quad i=1,2,3,4,9,10, \\
u^2\times 1.0245 &\quad i=5,7, \\
u^2\times 0.9006 &\quad i=6,8, \\
\end{array}\right. \\
\left(\dfrac{Z^g_{ij}}{(1-w_0^2)^2 Z_w^2}\right)^{\rm MF} &=& \left\{
\begin{array}{ll}
= Z^g_{ji}= 0.0416 &\quad (i,j) =(1,2),(3,4),(9,10), \\
-0.0412    &\quad (i,j) =(5,6),(7,8), \\
0 &\quad\mbox{others} \\
\end{array}\right. \\
\left(\dfrac{Z_i^{\rm pen}}{(1-w_0^2)^2 Z_w^2}\right)^{\rm MF} 
&=& C(Q_i)\times 0.0130, 
\een
at $\mu a = 1 $ in the DRED scheme, where we factor out 
the tadpole contributions.
We find that the penguin diagram 
contributions to the renormalization factor are quite small.

\section{Conclusion}
\label{sec:conclusion}  

In this paper we have presented the full expressions for the 
renormalization factors of the 
$\Delta S=1$ four-quark operators up to the one-loop level 
including the contributions of both the gluon exchange and penguin
diagrams. Our calculation, however,  does not include mixing with lower
dimensional operators ${\bar s}d$ and ${\bar s}\gamma_5 d$.
The coefficients of these operators
diverge as inverse powers of the lattice spacing, due
to which it is practically inappropriate 
to subtract these lower dimensional operators
by the perturbation theory. 
We are instead forced to use the non-perturbative methods such as those
based on the chiral perturbation theory\cite{np_chpt} and the Schr{\" o}dinger
functional scheme\cite{np_sf}. We leave it to future work. 

\section*{Acknowledgments}
This work is supported in part by the Grants-in-Aid for
Scientific Research from the Ministry of Education, Science and Culture
(Nos. 12640253, 12014202).
Y. K. is supported by Japan Society for Promotion of Science.

\section*{Appendix}                                                             

In this appendix we explain the domain-wall fermion action and its
Feynman rules relevant for the present calculation. 
The domain-wall fermion action is written as\cite{shamir},
\ben
S_{\rm DW} &=&
\sum_{n} \sum_{s=1}^{N_s} \Biggl[ \frac{1}{2} \sum_\mu
\left( \bpsi(n)_s (-r+\gamma_\mu) U_\mu(n) \psi(n+{\hat \mu})_s
+ \bpsi(n)_s (-r-\gamma_\mu) U_\mu^\dagger(n-{\hat \mu}) 
\psi(n-{\hat \mu})_s \right)
\nn\\&&
+ \frac{1}{2}
\left( \bpsi(n)_s (1+\gamma_5) \psi(n)_{s+1}
+ \bpsi(n)_s (1-\gamma_5) \psi(n)_{s-1} \right)
+ (M-1+4r) \bpsi(n)_s \psi(n)_s \Biggr]
\nn\\&+&
 m \sum_n \left( \bpsi(n)_{N_s} P_{R} \psi(n)_{1}
+ \bpsi(n)_{1} P_{L} \psi(n)_{N_s} \right),
\label{eq:action}
\een
where $U_\mu$ is the link variable of the SU($N$) gauge group 
and the Wilson parameter is set to $r=-1$.
Four dimensional space-time coordinate is labeled by $n$ and $s$ is an extra
fifth dimensional index which runs from $1$ to $N_s$.
Since we impose no gauge interaction along 
the fifth dimension, it is also possible to consider that the index $s$ labels
the $N_s$ ``flavor'' space.
In our one-loop calculation we take $N_s\to\infty$ to avoid
complications arising from the finite $N_s$
such as mixing among the four-quark operators with different chiralities.
The parameter $m$ denotes the physical quark mass and at $m=0$ 
one chiral zero mode is supposed to appear under the condition
$0 < M < 2$ for the Dirac ``mass'' $M$. 
$P_{R/L}$ are projection operators defined by $P_{R/L}=(1\pm\gamma_5)/2$.
For the gauge part we employ a standard four dimensional
Wilson plaquette action.

The quark fields on the four-dimensional space-time are given by the combinations
of the fermion fields at the boundaries,
\ben
q(n) = P_R \psi(n)_1 + P_L \psi(n)_{N_s},\\
\ovl{q}(n) = \bpsi(n)_{N_s} P_R + \bpsi(n)_1 P_L.
\label{eq:quark}
\een
We consider the composite operators constructed with these ``physical''
quark fields, and
our renormalization procedure is based on the Green functions
consisting of only these quark fields. 

In order to obtain the Feynman rules we perform the weak coupling expansion 
of the quark and gauge actions.
The fermion propagator with momentum $p$ 
is obtained by inverting the domain-wall Dirac operator
in eq.(\ref{eq:action}), which is expressed by $S_F(p)_{st}$ as 
an $N_s\times N_s$ matrix in the ``flavor'' space.
In the present one-loop calculation, however, we do not
need the whole matrix elements because
we consider the Green functions consisting of the
physical quark fields.
The relevant fermion propagators are restricted to
following three types:
\ben
\vev{q(-p) \ovl{q}(p)}&=& P_R S_F(p)_{11} P_L + P_L S_F(p)_{N_s N_s} P_R
+ P_L S_F(p)_{N_s 1} P_L + P_R S_F(p)_{1 N_s} P_R \nn\\
&=& 
 \frac{-i\gamma_\mu \sin p_\mu + \left(1-W e^{-\alpha}\right) m}
{-\left(1-e^{\alpha}W\right) + m^2 (1-W e^{-\alpha})},
\label{eqn:phys-prop}
\\
\vev{q(-p) \bpsi_s(p)}&=& P_R S_F(p)_{1s} + P_L S_F(p)_{N_s s} \nn\\
&=&
\frac{1}{F}
\left( i\gamma_\mu \sin p_\mu - m \left(1 -W e^{-\alpha} \right)
\right)
\left( e^{-\alpha (N_s-s)} P_R + e^{-\alpha (s-1)} P_L \right)
\nn\\&&
+\frac{1}{F} \Bigl[
m \left(i\gamma_\mu \sin p_\mu  -m \left(1-W e^{-\alpha}\right)\right)
- F \Bigr] e^{-\alpha}
\left( e^{-\alpha (s-1)} P_R + e^{-\alpha (N_s-s)} P_L \right),
\\
\vev{\psi_s(-p) \ovl{q}(p)} &=& S_F(p)_{s1} P_L + S_F(p)_{s N_s} P_R \nn \\
&=&
\frac{1}{F}
\left( e^{-\alpha (N_s-s)} P_L + e^{-\alpha (s-1)} P_R \right)
\left( i\gamma_\mu \sin p_\mu - m \left(1 - W e^{-\alpha} \right)
\right)
\nn\\&&
+\frac{1}{F}
\left( e^{-\alpha (s-1)} P_L + e^{-\alpha (N_s-s)} P_R \right) e^{-\alpha}
\Bigl[
m \left(i\gamma_\mu \sin p_\mu  -m\left(1- We^{-\alpha}\right) \right)
- F \Bigr]
\end{eqnarray}
with
\begin{eqnarray}
W &=& 1-M -r \sum_\mu (1-\cos p_\mu),
\\
\cosh (\alpha) &=& \frac{1+W^2+\sum_\mu \sin^2 p_\mu}{2|W|},
\label{eq:alpha}
\\
F &=& 1-e^{\alpha} W-m^2 \left(1-W e^{-\alpha}\right),
\label{eq:F}
\een
where the argument $p$ in the factors $\alpha$ and $W$ is
suppressed. 

In the perturbative calculation of Green functions 
we assume that the external quark momenta and masses are  
much smaller than the lattice cut-off. In this case
the external quark propagators can be expanded in terms of them.
We have the following expressions as the leading term of the expansion:
\ben
\langle q\bar q \rangle (p) & = & \frac{1-w_0^2}{i\pslash + (1-w_0^2)m},
\\
\langle q \bar \psi_s \rangle (p) &=&
\langle q\bar q \rangle (p)
\left( w_0^{s-1}P_L + w_0^{N_s-s} P_R\right),
\label{eqn:qpsi}
\\
\langle \psi_s \bar q\rangle (p) &=&
\left( w_0^{s-1}P_R + w_0^{N_s-s} P_L\right) 
\langle q\bar q \rangle (p),
\label{eq:psiq}
\een
where $w_0 = 1-M$. The form of 
$\langle q\bar q \rangle (p)$  tells us that $\sqrt{1-w_0^2}$ is 
the overlap factor to the four-dimensional quark fields at the tree-level.

The one-gluon fermion vertex with outgoing and incoming fermion momenta  
$p$ and $q$, respectively, is given by
\ben
V_{1\mu}^A (q,p)_{st}=g T^A v_\mu\left(\frac{q_\mu+p_\mu}{2}\right)_{st}
&=& g T^A \{ i\gamma_\mu \cos\left(\frac{q_\mu+p_\mu}{2}\right)
  + r \sin\left(\frac{q_\mu+p_\mu}{2}\right) \} \delta_{st},
\een
where $T^A$ $(A=1,\dots,N^2-1)$ are generators of color 
SU($N$). We do not need two-gluon fermion vertex in the present calculation. 

The gluon propagator for a gluon of momentum $p$ is written as 
\ben
G_{\mu \nu}^{AB} (p)
=\frac{\delta_{AB}}{4\sin^2 (p/2)}
\left[\delta_{\mu \nu}
- (1-\alpha) \frac{4 \sin ({p}_\mu/2) \sin ({p}_\nu/2)}{4 \sin^2 (p/2)}
\right],
\een                                                                  
where $\sin^2(p/2) =\sum_\mu \sin^2(p_\mu/2)$ and we choose the Feynman gauge
($\alpha=1$) in our calculation.

\newcommand{\J}[4]{{\it #1} {\bf #2} (19#3) #4}
\newcommand{\MPL}{Mod.~Phys.~Lett.}
\newcommand{\IJMP}{Int.~J.~Mod.~Phys.}
\newcommand{\NP}{Nucl.~Phys.}
\newcommand{\PL}{Phys.~Lett.}
\newcommand{\PR}{Phys.~Rev.}
\newcommand{\PRL}{Phys.~Rev.~Lett.}
\newcommand{\AP}{Ann.~Phys.}
\newcommand{\CMP}{Commun.~Math.~Phys.}
\newcommand{\PTP}{Prog. Theor. Phys.}
\newcommand{\Suppl}{Prog. Theor. Phys. Suppl.}

\tightenlines
\begin{table}
\caption{Numerical values for $A$, $B$ and 
$z_i^{\rm pen}$(DRED) as a function of $M$. Note that $z_i^{\rm pen}$ 
is independent
of $i$ ($i=1,\dots,10$) in the DRED scheme.}
\label{tab:penguin}
\begin{center}
\begin{tabular}{lll|l}
$M$ & $A$ & $B$ & $z_i^{\rm pen}$(DRED) \\
\hline
0.05 & -1.5078(4) &    1.5077(3) &      5.729(1) \\
0.10 & -0.1130(4) &    0.1128(3) &      1.544(1) \\
0.15 &  0.2836(4) &   -0.2838(3) &      0.355(1) \\
0.20 &  0.4456(4) &   -0.4458(3) &     -0.131(1) \\
0.25 &  0.5193(4) &   -0.5195(3) &     -0.352(1) \\
0.30 &  0.5513(4) &   -0.5515(3) &     -0.448(1) \\
0.35 &  0.5607(4) &   -0.5609(3) &     -0.476(1) \\
0.40 &  0.5565(4) &   -0.5567(3) &     -0.464(1) \\
0.45 &  0.5434(4) &   -0.5436(3) &     -0.425(1) \\
0.50 &  0.5242(4) &   -0.5244(3) &     -0.367(1) \\
0.55 &  0.5003(4) &   -0.5006(3) &     -0.296(1) \\
0.60 &  0.4728(4) &   -0.4729(3) &     -0.213(1) \\
0.65 &  0.4419(4) &   -0.4421(3) &     -0.120(1) \\
0.70 &  0.4081(4) &   -0.4082(3) &     -0.019(1) \\
0.75 &  0.3712(4) &   -0.3714(3) &      0.092(1) \\
0.80 &  0.3314(4) &   -0.3316(4) &      0.211(1) \\
0.85 &  0.2884(4) &   -0.2885(3) &      0.341(1) \\
0.90 &  0.2417(4) &   -0.2419(3) &      0.480(1) \\
0.95 &  0.1911(4) &   -0.1913(3) &      0.632(1) \\
1.00 &  0.1360(4) &   -0.1362(4) &      0.797(1) \\
1.05 &  0.0760(11)&   -0.0754(4) &      0.980(1) \\
1.10 &  0.00879(7)&   -0.00944(65) &    1.178(2) \\
1.15 & -0.0648(3) &    0.0647(4) &      1.400(1) \\
1.20 & -0.1476(3) &    0.1472(4) &      1.648(1) \\
1.25 & -0.2407(4) &    0.2405(3) &      1.928(1) \\
1.30 & -0.3470(4) &    0.3468(4) &      2.246(1) \\
1.35 & -0.4695(4) &    0.4693(4) &      2.614(1) \\
1.40 & -0.6125(4) &    0.6122(3) &      3.043(1) \\
1.45 & -0.7818(4) &    0.7816(4) &      3.551(1) \\
1.50 & -0.9859(4) &    0.9857(3) &      4.163(1) \\
1.55 & -1.2369(4) &    1.2367(3) &      4.916(1) \\
1.60 & -1.5531(4) &    1.5529(3) &      5.865(1) \\
1.65 & -1.9638(4) &    1.9636(3) &      7.097(1) \\
1.70 & -2.5179(4) &    2.5177(3) &      8.759(1) \\
1.75 & -3.3049(4) &    3.3047(4) &     11.120(1) \\
1.80 & -4.5053(4) &    4.5051(3) &     14.721(1) \\
1.85 & -6.5451(3) &    6.5449(3) &     20.841(1) \\
1.90 &-10.7173(1) &   10.717(3)  &     33.357(1) \\
1.95 &-23.549(7)  &   23.549(7)  &    71.853(22) \\
\end{tabular}
\end{center}
\end{table}

\begin{figure}[t]
\centering{
\hskip -0.0cm
\psfig{file=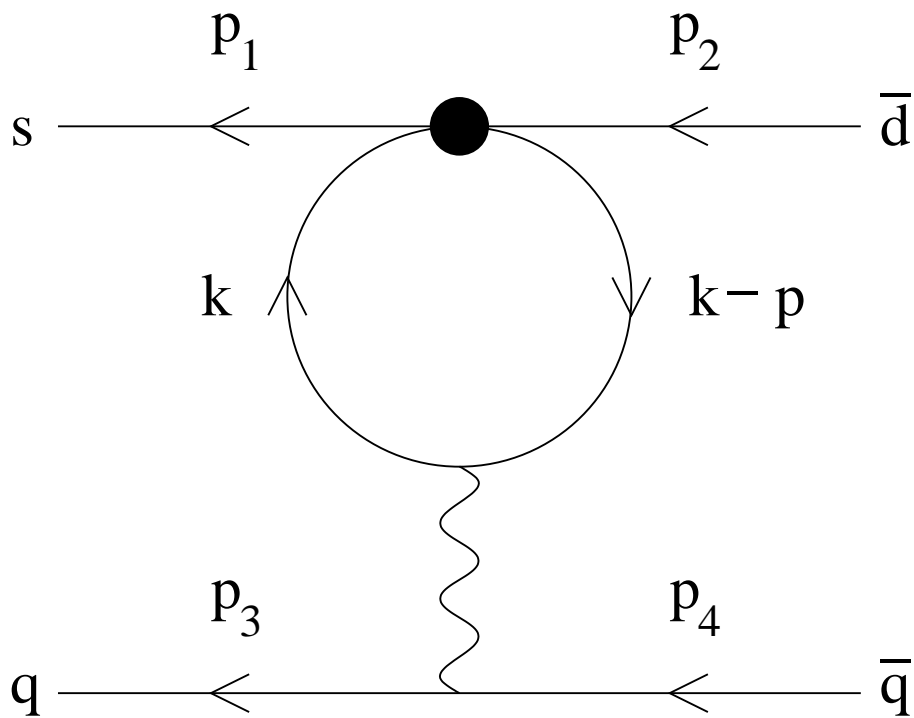,width=120mm,angle=-90}     
\vspace{5mm}  }
\caption{The penguin diagram. $s$, $d$ and $q$ label
quark flavors. $p_1$, $p_2$, $p_3$, $p_4$, $p$ and $k$ 
are momenta. Solid circle denotes the four-quark operator.}
\label{fig:penguin}
\vspace{8mm}
\end{figure}                                                                    

\end{document}